# Optical Measurement of Pseudo-Spin Texture of the Exciton Fine-Structure in Monolayer WSe$_2$ within the Light Cone

Lorenz Maximilian Schneider[1], Shanece S. Esdaille[2], Daniel A. Rhodes[2], Katayun Barmak[3], James C. Hone[2,*], and Arash Rahimi-Iman[1,*]

[1]*Faculty of Physics and Materials Sciences Center, Philipps-Universität Marburg, Marburg, 35032, Germany*

[2]*Department of Mechanical Engineering, Columbia University, New York, NY 10027, USA*

[3]*Department of Applied Physics and Applied Mathematics, Columbia University, New York, NY 10027, USA*

**Abstract:**

Several theoretical predictions have claimed that the neutral exciton of TMDCs splits into a transversal and longitudinal exciton branch, with the longitudinal one, which is the upper branch, exhibiting an extraordinary strong dispersion in the meV range within the light cone. Historically, this was linked for semiconductor quantum wells to strong far-field optical dipole coupling, or strong electronic long-range exchange interactions, describing two sides of the same coin. Recently, experiments utilizing Fourier-space spectroscopy have shown that the exciton (exciton–polariton) dispersion can indeed be measured for high-quality hexagonal-BN-encapsulated WSe$_2$ monolayer samples and can confirm the energy scale. Here, the exciton fine-structure's pseudo-spin and the valley polarization are investigated as a function of the centre-of-mass-momentum and excitation-laser detuning. For quasi-resonant excitation, a strong dispersion featuring a pronounced momentum-dependent helicity is observed. By increasing the excitation energy step-wise towards and then above the electronic band gap, the dispersion and the helicity systematically decrease due to contributions of incoherent excitons and emission from plasma. The decline of the helicity with centre-of-mass momentum can be phenomenologically modelled by the Maialle–Silva–Sham mechanism using the exciton splitting as the source of an effective magnetic field.

## Introduction

2D-materials studies have made tremendous progress in the last years with regard to electronic and optical characterization of basic properties, especially based on the now commonly-used encapsulation in hexagonal boron nitride (h-BN) [1–6], which increases the optical quality of the samples dramatically by decreasing the charge disorder. While this has opened the path to the exploration of a variety of different application scenarios [7–10], it has also changed the charge-carrier screening important to many fundamental investigations.

Theoretical work has shown that the carrier screening in these 2D semiconductors and metals is indeed very different from that of their 3D counterparts. The renormalization of Coulomb energies is not scale invariant [11], but the tiny height of the material introduces a characteristic length scale; or if speaking in terms of the reciprocal space, the dielectric function is (linearly) dependent on the momentum. This means that for the case of a suspended layer [11] for small distances the screening creates effectively a Rytova-Keldysh potential [12,13], while for distances between hole and electron larger than the characteristic length of that system, the effective potential behaves like the unscreened 3D-Coloumb potential. This change of screening could be proven by the observation of a non-hydrogenic Rydberg series [14]. Caused by this reduced screening, the monolayered material features comparably strong Coulomb interactions especially for long range interactions, and it exhibits a large exciton binding energy. It is clear that when the monolayer is transferred to a substrate or encapsulated with a dielectric such as h-BN, changes to the screening model [15–18] have to be applied owing to the induction of mirror charges in the environment. Nonetheless, the fundamental difference between the 2D material and the 3D crystals in terms of a momentum-dependent screening stays valid even when encapsulated as long as the permittivity contrast between the 2D material and the surrounding is significant.

The reduced screening enhances not only the effects by the direct Coulomb term but also by those related to its exchange interaction. Therefore, Yu et al. [19] and other scientists [20–23] rediscovered the concept of an enhanced or even linear dispersion of the longitudinal exciton branch caused by the long-range exchange interactions already proposed for quantum-well (exciton–)polaritons in III-V semiconductors [24,25]. While in III/V quantum wells the effect is negligible and smaller than the usual quadratic dispersion, in 2D-materials, where the interaction strength is about two orders of magnitude stronger, it has been predicted to be on the order of 1.5 meV within the light cone. Here, Yu et al. [19], Wu et al. [20] and Qiu et al. [21] predicted a linear slope of the dispersion relation around zero centre-of-mass momentum, as they use a strictly 2D-Coloumb potential, whereas Deilmann et al. [23] claimed merely an increased curvature of the underlying parabolic dispersion by using a 3D-Coloumb potential. The authors argued that for realistic samples on substrate or encapsulated monolayers a strict 2D treatment is not valid anymore.

Furthermore, Refs. [19–21] reported that the longitudinal (upper) and transversal (lower) branch have to be considered as coherent superpositions of both valleys, with the energetic separation between the hybridized states resembling the Rabi oscillations between the Rabi-split states. Thus, the valley pseudo-spin is necessarily linked with the centre-of-mass motion and a measurement of the splitting can give insights into the dephasing mechanism of the valley-pseudo spin by exchange interactions or decoherence.

Recently, the WSe$_2$ exciton (exciton–polariton) dispersion within the light cone was measured successfully both in white-light reflection and in photoluminescence by optical Fourier-space spectroscopy [6] on high-quality h-BN encapsulated samples. Indeed, a measurable dispersion in the meV range could be experimentally demonstrated, giving access to intriguing studies of light–matter interactions and exciton physics in 2D materials. To leap one crucial step further, here, we investigate the phase-space-dependent pseudo-spin texture of the rich exciton fine-structure in such encapsulated WSe$_2$ monolayer by optical helicity measurements and reveal the polarization- and excitation-energy dependence of the neutral exciton (exciton–polariton) dispersion branch as a function of the in-plane momentum. A systematic decrease is evidenced due to contributions of incoherent excitons and emission from plasma when moving the excitation energy step-wise above the electronic band gap. Our findings indicate—in accordance with the strong dispersion feature and valley-polarization degree—an ultralight species of delocalized and hybridized quasi-particles in the valley (pseudo-spin) landscape of the WSe$_2$ monolayer's 2D lattice. Remarkably, high-quality 2D stacks offer these highly-mobile correlated electron–hole pairs with pronounced helicity, strong in-plane long-range dipole interactions and macroscopic polarization.

**Experiment**

Generally, when a light beam is incident onto a monolayer under an angle with respect to the plane's normal, the electric field (polarization) of the light wave can be decomposed into a longitudinal (oriented in the plane of incidence parallel to the monolayer), a transversal (orthogonal to the propagation direction and in the plane of the monolayer) and an out-of-plane component (cf. **Figure 1a**). These different field components couple to the different branches of the exciton fine structure. The longitudinal component couples to the upper branch of the bright exciton with non-analytic dispersion, the transversal part to the lower branch and the out-of-plane projection to the z-mode, which is often also called a dark exciton [20–22,26,27] (cf. schematic in **Figure 1b**). The underlying parabolic free-exciton dispersion of all three species is in theory only strongly modified by the coupling to the longitudinal polarization in the case of the upper bright branch, rendering it indistinguishable from the lower bright branch at exact zero momentum (normal incidence) due to the maximum induced energy shift (the two polarizations in the plane become indistinguishable at α = 0°). The modified dispersion recovers to the uncoupled curve at angles approaching 90°, corresponding to a vanishing longitudinal field component.

To analyze the valley pseudo-spin of these branches, polarization-resolved PL spectra with $k_{||}$-space resolution have been acquired using different excitation energies under continuous-wave (cw) pumping. The sample is a high-quality h-BN-encapsulated WSe$_2$ ML employed in a cryostat and measured optically using a µ-PL setup with a 4f geometry [6]. The used excitation energies are 1.784 eV, 1.810 eV, 1.952 eV and 2.330 eV and cover distinct detunings from quasi-resonant (A$_{1s}$ as well as $A_{2s}$) to continuum excitation (above the electronic band gap, where an unbound correlated electron–hole pair resembles the A exciton in its ionized state). The excitation power was kept constant at 0.4 mW for all situations. Note at the same time by changing the energy the photon flux will change by 30%. Nonetheless, all situations are in the linear regime resulting in an approximate carrier density of about $10^9 cm^{-2}$. A schematic diagram indicating the relative pump levels with respect to the excitonic states in WSe₂ according to Ref. [21,28] is shown in **Figure 1c**.

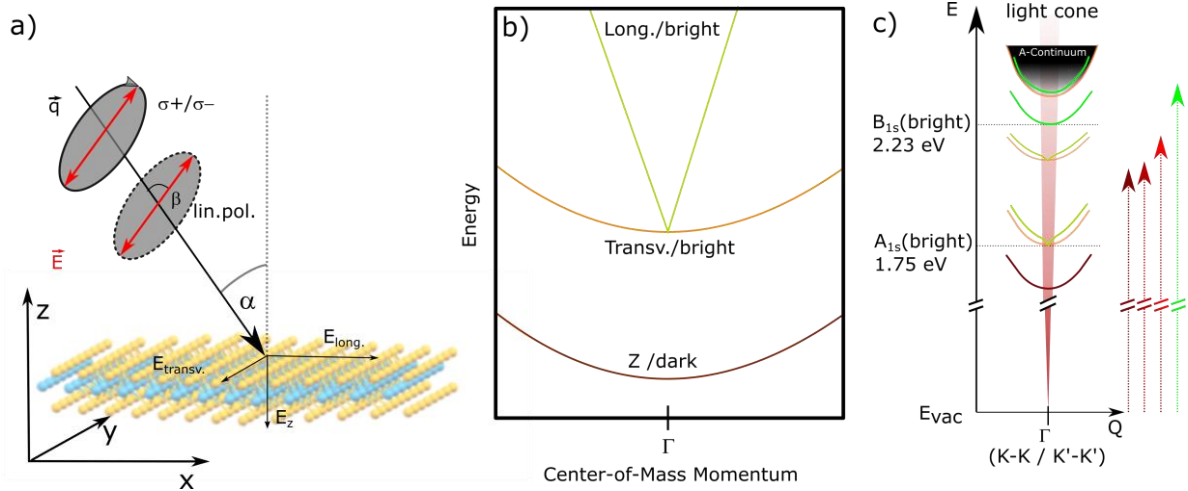

**Figure 1 | Sketch of the sample excitation and exciton fine-structure dispersions.** (a) Schematic of the incident radiation on the ML WSe₂ with its electric-field polarization components in relation to the incidence as well as the sample plane (angle α and grey-shaded polarization-rotation circle, respectively). Thus, every polarization splits into a field component in longitudinal as well as out-of-plane direction, and a component in transverse orientation. (b) Magnified sketch of the predicted energy-momentum dispersion for longitudinal and transversal excitons with degeneracy at zero in-plane momentum, as well as the dark exciton (also referred to as z-mode) which corresponds to an out-of-plane dipole. (c) Schematic dispersion overview diagram in the two-particle picture for the monolayer excitons, indicating the energies of the A and B bright excitons (according to Ref. [21,28]) together with the used excitation energies.

The experimentally obtained Fourier-space-resolved spectra (energy vs. in-plane-momentum plot; short: far-field [FF] spectrum) at 10 K are shown in **Figure 2a–2d** and have been recorded without selecting a particular polarization of the emitted light. Momentum-integrated line spectra are displayed onto the FF spectra for comparison. Remarkably, a curved—i.e. measurable—dispersion of the neutral exciton can be measured via FF spectroscopy for quasi-resonant pumping on the meV scale as known from [6]. For an overview reflection contrast FF spectrum, which also exhibits a visibly curved dispersion (a noticeable one for the 1s and a subtle one for the 2s resonance), see the Supporting Information. In contrast to differently-detuned quasi-resonant excitations of the A exciton, only a flat line is obtained for off-resonant pumping, i.e. a non-resolvable dispersion is measured. The former case, i.e. the behavior obtained for quasi-resonant excitation, is expected by the discussed theories.

In contrast, for off-resonant excitation, the coherence times are expected to be very low. Thus, macroscopic polarization is quickly transformed into microscopic polarization due to decoherence. The long-range interaction, which in a simplified picture can be understood as the backaction (feedback) of the electric field from the induced polarization—induced by the dipole of the exciton— on the exciton itself [24,25,29], is therefore lower due to the loss of a macroscopic dipole when compared to a case with long coherence time. In this picture, such long-range interaction between excitations in different valleys is very similar to the formation of exciton–polaritons — i.e. dipole–dipole coupling via a macroscopic polarization takes place. In the case of weak interaction at off-resonant pumping, the separate valleys can be still seen as the (uncoupled) eigenstates of the system and no hybridization occurs. However, in the other extreme, a purely coherent long-lasting excitation in both valleys would strongly interact and hybridize the eigenstates.

In between these two cases, when pumping the 2D crystal shortly above its 2s resonance (**Figure 2c**) of the neutral exciton, it seems as if the observation exhibits a mixture of, both, a strong dispersion from coherent, hybridized valley states, as well as a flat dispersion arising from non-coherent contributions. The observation of a vanishing curvature can be understood as a decreasing fraction of excitons. Full many-body calculation for GaAs quantum wells [30] showed that at increased detuning (even if just above the 2s-bound-state resonance) the population fraction of excitons in the 1s state drops dramatically, namely from about 80% at a resonant condition to 20% when pumping the 2s state resonantly, and to only 3% when pumping the system above the quasi-particle band gap, while the rest corresponds to an uncorrelated electron–hole plasma. Our observations are in qualitative agreement with these calculations.

The extracted dispersion relations from **Figure 2a** to **2d** are shown in **Figure 2e**. Although a number of theories predict a linear slope, we have followed the argument of Deilmann et al. [23] and extracted the effective mass by parabolic fits to the extracted position of the neutral exciton as it describes the data well and allows for easy comparison with polariton systems. The bright exciton's average effective mass gets heavier starting from $1\times10^{-4}$ $m_0$ at quasi-resonant pumping to more than $1\times10^{-3}$ $m_0$ with increased detuning until finally no parabolic fit is applicable when the system is pumped into the continuum. Note that the change of effective mass is attributed to the mixture of coherent and incoherent components in the FF spectra which affect the extraction of dispersion curves, whereas the underlying curvatures of the different emitters is not necessarily altered. The small tilt in **Figure 2e** visible for the actually flat dispersion is caused by a slightly tilted orientation of the imaging monochromator's CCD camera (rotation of 0.5°).

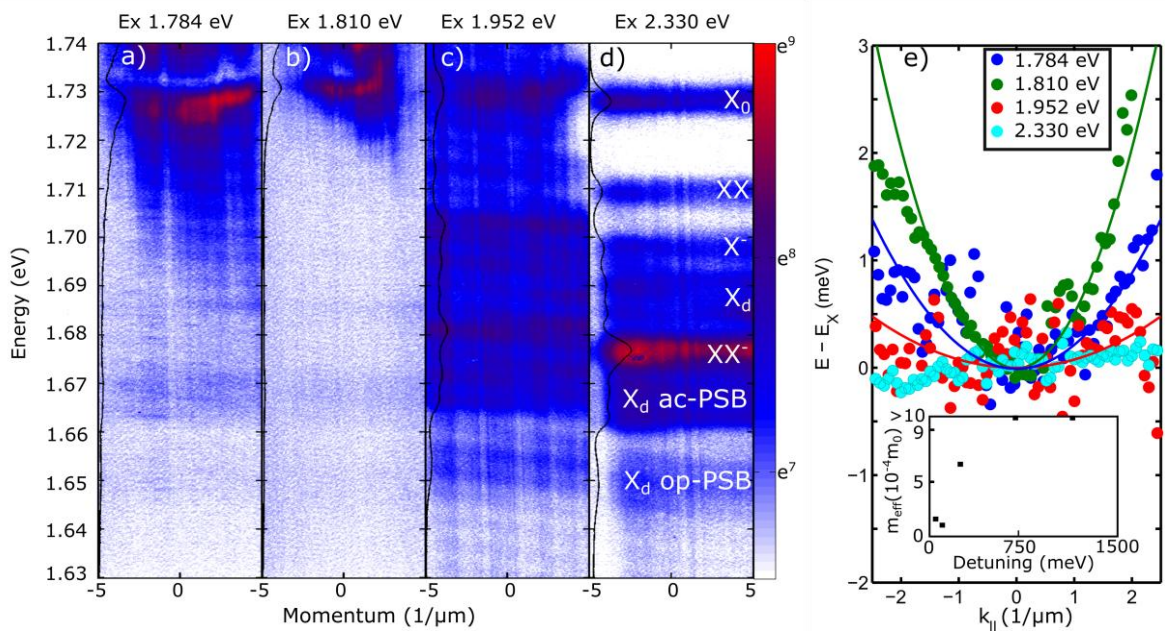

*Figure 2* | **PL Fourier-space-resolved spectra for different excitation detunings.** (a-d) Logarithmic false-colour contour plots of the total intensity under right-circular-polarized pumping for different excitation energies (labelled Ex.) and fixed low photon flux at 10 K. The signal was not detected with polarization sensitivity. The line spectra displayed as insets represent angle-integrated data. The different excitonic features are labelled in (d) similar to Refs. [31–35]. (e) Neutral exciton's extracted peak positions as a function of the in-plane momentum for average effective mass estimations using parabolic dispersion fits.

Interestingly, while for quasi-resonant pumping, the spectrum indicates equidistantly separated replicas of the exciton mode attributed to phonon sidebands, the higher the chosen pump energy, the more exciton complexes become visible in the spectrum. When comparing the spectra, one can see that the emission from trions and biexcitons increases significantly when the excitation energy is raised above the energy of the 2s state. While the emission from phonon sidebands of dark exciton states [31,35–37] from 1.65 eV to 1.67 eV and from z-mode excitons [26,27,31] around 1.69 eV is weaker at quasi-resonant conditions, it can be measured at all excitation conditions studied here.

Ultimately, to study the helicity, i.e. the degree of circular polarization, for the different excitation conditions, FF spectra with a selected circular polarization under circularly polarized pumping have been acquired. The amount of helicity is calculated using the common formula $\rho = (I^{\sigma+} - I^{\sigma-})/(I^{\sigma+} + I^{\sigma-})$ and is plotted in **Figure 3a–3d.** For the quasi-resonant 1s excitation cases, a pronounced helicity of about 40% is obtained for the neutral exciton. When exciting above the 2s resonance, the helicity is reduced but still clearly visible in comparison to the case when the system is excited above its quasi-particle band, where it is almost absent similar to previous characterizations without angle resolution [38–40]. Remarkably, these helicity maps for FF spectra all indicate the considerable amount of the circular-polarization degree for various exciton complexes as expected, clearly highlighting the valley-excitonic nature (in contrast to defect state emission) with the valley-selective excitation and detection scheme. For clarity, these four false-colour contour maps share the same helicity scale and are, thus, directly comparable.

In order to analyze the momentum dependence of the helicity for the neutral exciton species, the maximum measured helicity value in the relevant energy range of 1.725 eV and 1.733 eV has been extracted along the momentum axis for all different excitation energies (see **Figure 4a**). For the quasi-resonant excitation cases, one can see a drop of helicity with increasing in-plane momentum. In the intermediate case only a small variation in helicity is obtained, which is only slightly bigger than the noise. However, in the case of continuum excitation, no correlation between the centre-of-mass momentum and helicity can be found, as expected. As Yu et al. [19] explained that the splitting between the states of the bright exciton resembles their Rabi oscillation, it is not surprising that a smaller helicity is evidenced at larger centre-of-mass momentum and correspondingly larger energetic splitting between the branches, as a stronger interaction will lead to a faster depolarization.

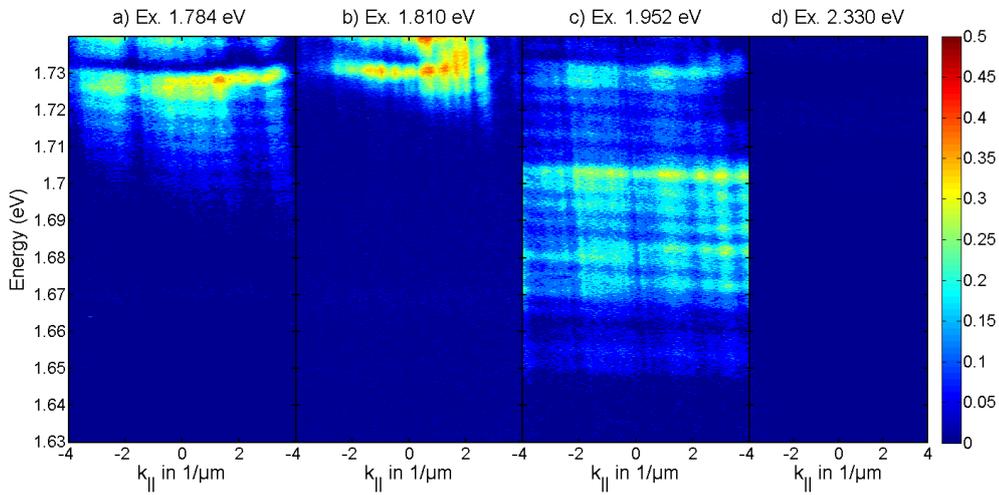

**Figure 3 | PL helicity analyzed utilizing Fourier-space spectra at different detuning of excitation.** (a-d) Fourier-space maps of the emitters' helicity obtained from polarization-selective measurements for different detuning (Ex.: excitation energies).

**Discussion**

The strong dispersion observed here has its origin in the lifting of the degeneracy between a longitudinal and transversal exciton(–polariton) branch due to long-range exchange interaction. The excitonic Hamiltonian reads [20,22,41]:

(1) $$H_k = \left(\hbar\omega_0 + \frac{\hbar^2 k^2}{2M} + J_k\right)\sigma_0 + J_k[\cos(2\phi_k)\,\sigma_x + \sin(2\phi_k)\,\sigma_y].$$

Here, $\boldsymbol{\sigma}$ are the Pauli matrices, $M$ the exciton mass, $J_k$ the additional energy due to the long-range exchange interaction and $\phi_k$ the orientational in-plane angle of the centre-of-mass momentum. In such a scenario, where excitons polarized in two perpendicular in-plane directions have different energy, it is established knowledge that this leads to spin and valley decoherence [22,42]. This is easily understood, as the interaction leads to a mixing of the pure valley states [22]. This spin-decoherence is normally modelled by introducing an effective magnetic field $\boldsymbol{\Omega}_{eff}(k_{||})$ and has been used to model classical semiconductors [43], angle-integrated helicity of excitons in 2D-monolayers [22] as well as the angle-resolved helicity found for valley cavity–polaritons based on TMDC [41,44]. The dynamics of the valley pseudo-spin $\mathbf{S}_k$ then follows the following equation:

(2) $$\frac{\partial S_k}{\partial t} = Q\{\mathbf{S}_k\} - \mathbf{S}_k \times \mathbf{\Omega}_{eff}(k).$$

Here, $Q$ is the collisional integral, which describes other decoherence pathways such as scattering with phonons and other excitons. In case of excitons in TMDCs, the effective magnetic field can be denoted as follows [20,21,41] with the excitonic Hopfield coefficient $|X_k|^2$ and the energetic splitting of the exciton branches $\Omega_{ex}(k)$:

(3) $$\mathbf{\Omega}_{eff}(k) = |X_k|^2 \mathbf{\Omega}_{ex}(k) = |X_k|^2 2 J_k (\cos(2\phi_k), \sin(2\phi_k), 0)/\hbar = |X_k|^2 2Ak(\cos(2\phi_k), \sin(2\phi_k), 0)/\hbar$$

The effective field vanishes at zero momentum. *A* denotes the strength of the in-plane interactions. Under the assumption that the collisional integral dependence on the momentum is negligible within the light cone, the resulting decoherence time $\tau_{deco}$ can be decomposed into a momentum-independent part $\tau_{deco,k=0}$ and the *k*-dependent decoherence part due to exchange interactions $\tau_{deco,EI}$. If one assumes further that the spin decoherence happens by the Maialle-Silva-Sham mechanism [45] similar to previous reports on cavity–polaritons in TMDCs [41,44], the spin-decoherence time can be written as follows with the momentum relaxation time $\tau_{relax}$ :

(4) $$\frac{1}{\tau_{deco(k)}} = \frac{1}{\tau_{deco,k=0}} + \frac{1}{\tau_{deco,EI(k)}} = \frac{1}{\tau_{deco,k=0}} + \Omega_{eff,\|}^2(k)\tau_{relax}.$$

Note that Glazov et al. [22] assumed a Dyakonov-Perel type mechanism instead. As in both cases motional narrowing results in the spin decay, this will not change equation **(4)**. As the long-range part does not contribute to the decoherence at zero momentum, the momentum-independent pseudo-spin decoherence time can be directly estimated using the following formula (after [22]):

(5) $$\rho_c = \rho_0/(1 + \frac{\tau_{rec}}{\tau_{deco,k=o}}).$$

Assuming strict selection rules ($\rho_0 = 100\%$), a recombination time of $\tau_{rec} = 5.0 \pm 0.5 \, ps$ similar to previous reports [22,41,4] and the experimentally measured degree of circular polarization $\rho_{C,max}$ (**Figure 4b**), one can determine the *k*-independent decoherence time (**Figure 4d**).

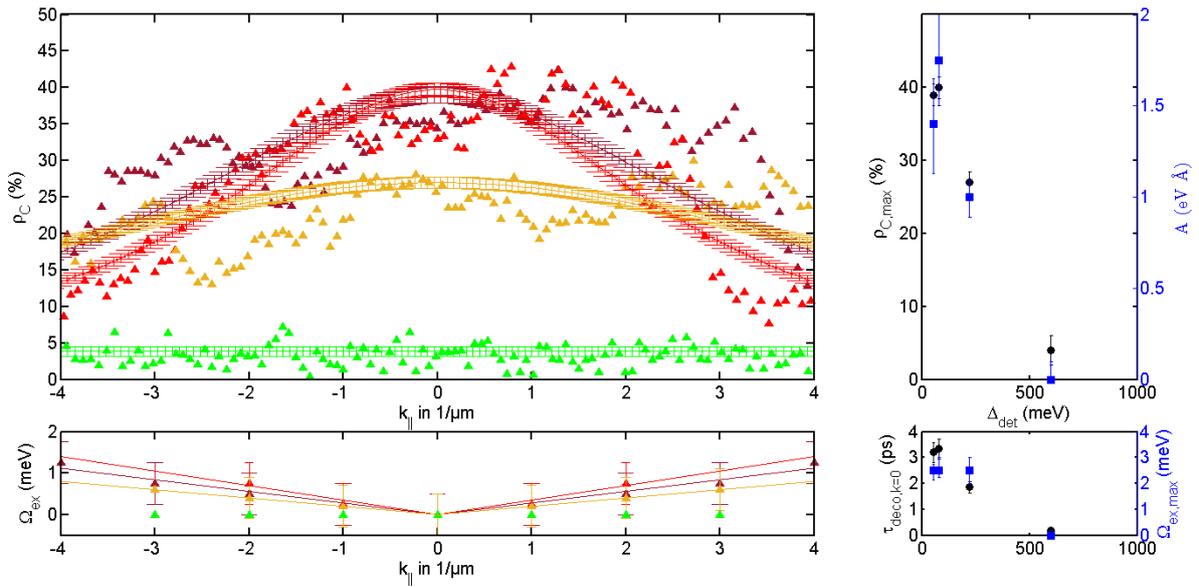

**Figure 4 | Comparison of the *k*-dependent PL helicity with the polarization model.** (a) Experimentally observed degree of helicity $\rho_c$ (triangles) in comparison with model predictions (colour-coded lines with error bars). The error bars denote to the calculated propagated uncertainty from the assumed or derived input parameters. (b) Plot of the maximum observed helicity (left axis) and the slope of the linear branch *A* as a function of detuning (right axis). The standard deviation is the approximated extraction error from (a) or the confidence interval of the linear fit in (c), respectively. (c) Manually extracted splitting of the branches from experimental Fourier-space PL spectra (using first derivative representation). The errors denote the approximated accuracy of extraction. (d) Phenomenologically determined valley decoherence time $\tau_{deco,k=0}$ and estimated general maximum splitting of the exciton branches within the light cone (a detuning-independent constant) used for modelling.

In order to extract the effective magnetic field, the energy derivative of the Fourier-space spectra were plotted and the approximated splitting for distinct *k* values manually was extracted (**Figure 4c**). An example derivative spectrum is shown in the Supporting Information. The resulting energy splitting was fitted with a line to extract the slope $2A$ of the linear branch (see **Figure 4b,c**). Note that, while the linear fits suggest a large splitting at the light cone edge, Qiu et al. [21] predicted a saturation towards the edge such that a maximum splitting of 2.0 to 2.7 meV at the edge can be expected from this experimental data. For the cases with measurable dispersion, the underlying plausible value of 2.5 meV is used (**Figure 4d**).

Finally, if a pure exciton gas ($|X_k|^2 = 1$) and a momentum relaxation time of 241 fs ($2\hbar/\Gamma$ using a linewidth of 6 meV) is assumed, the decoherence time can be calculated using equation **(4)** and the polarization using equation **(5)**. The resulting predictions for all detuning cases are shown together with the experimental helicity in **Figure 4a** which is in good agreement with the experimental observations.

The model shows that the decaying helicity is indeed most likely arising from the *k*-dependent long-range exchange interaction. Similarly, the valley decoherence due to other processes is strongly enhanced at increased detuning, as expected.

**Conclusion**

To summarize, we observe a dispersion of the neutral exciton emission under quasi-resonant circularly-polarized cw excitation conditions attributed to coherent mixing of the excitonic states at the **K** and **K'** valley in qualitative agreement with the predictions for monolayer TMDCs [19–22,24]. Remarkably, but not surprisingly, the curved dispersion is gradually vanishing if the excitation detuning is increased with respect to the neutral exciton resonance. This is understood when taking into account a faster decoherence of the coherently mixed exciton states with increasing detuning. In the case of quasi-resonant excitation, where strong coherence is expected, a decline of the mode's helicity with increasing angle is observed compared to normal incidence. A phenomenological model based on the Maialle-Silva-Sham mechanism [45] on the basis of the extracted energy splitting shows good agreement with the observations. Accordingly, the increased long-range exchange interaction towards higher in-plane momenta facilitates depolarization. This effect also reduces with increased detuning, for instance as noticeable for pumping the 2s state quasi-resonantly. However, in the case of continuum excitation, where basically only plasma and an incoherent exciton population occur [29], no correlation between centre-of-mass momentum and helicity is found, as expected. Revealing that there is a complex locking between valley-pseudospin and centre-of-mass momentum present, our study encourages further investigations of angle-resolved polarization anisotropy in exciton-complexes-rich 2D semiconductors. In particular, time-resolved Fourier-space mapping may further shed light on valley decoherence mechanisms as function of the centre-of-mass momentum.

**Methods**

**Sample preparation.** Tungsten diselenide (WSe$_2$) bulk single crystals were grown in an excess selenium flux (defect density: 5 x 10$^{10}$/cm$^2$) (see Ref. [46]). For encapsulated samples, monolayer WSe$_2$ and h-BN were first exfoliated from bulk single crystals onto SiO$_2$. For WSe$_2$, the SiO$_2$ substrate was first exposed to an O$_2$ plasma step before exfoliation. Monolayers and thin h-BN were both identified by optical contrast using a microscope. Afterwards, a dry stacking technique using polypropylene carbonate (PPC) on PDMS was used to pickup and stack h-BN/WSe$_2$ layers. First a top layer of h-BN is picked up at 48 degrees C, then WSe$_2$, and finally the bottom layer of h-BN. After each h-BN pickup step the PPC is briefly heated to 90 degrees C to re-smooth the PPC and ensure a clean wave front. For transferring the stack onto a clean substrate, the substrate is first heated to 75 degrees C, the stack is then put into contact, and gradually heated to 120 degrees C. Afterwards, the PPC/PDMS is lifted and the substrate is immersed in chloroform and rinsed with IPA to remove polymer residue. Atomic-force microscopy confirmed a total stack thickness of ~40 nm (~10 nm + ~30 nm for the encapsulating top and bottom h-BN, respectively).

**Optical measurement.** Angle-resolved micro-photoluminescence (µ-PL) measurements were performed using a self-built confocal optical microscope with Fourier-space imaging capabilities (similar to Ref. [6]). The sample was mounted in an evacuated (~10$^{-7}$ mbar) helium-flow cryostat, which was placed under the 40x microscope objective of the µ-PL setup. All data shown are time-integrated spectra. For quasi-resonant excitation, a continuous-wave-operated titanium-sapphire laser (*SpectraPhysics Tsunami*) was used, emitting light at about 695 or 685 nm. A high aspect-ratio long-pass filter with specified edge at 700 nm was used to block the laser light. Also, higher-detuning pumping was achieved with continuous-wave lasers. For detection, a nitrogen-cooled charge-coupled device (CCD) behind an imaging monochromator (*Princeton Instruments Acton SP2300*) was used, using two-dimensional chip read-out. Dispersion curves that are intensity (counts per integration time) over wavelength (long axis) over momentum (short axis) were recorded via the exposed CCD area. For *k*-space-resolved spectroscopy, the Fourier-space image of the sample's emission is projected onto the monochromator entrance slit via a set of lenses. The light cone in the medium (vacuum) amounts to $|k_{||}|$~ 34 (8.7) µm$^{-1}$, the maximum detectable angle of ±37° corresponds to $|k_{||}|$~5.2 µm$^{-1}$ in WSe$_2$. Relative effective-mass fit certainty is about 5 – 10 %. Detection of the µ-PL signal from the sample took place behind a spatially-filtering aperture in the real-space projection plane of the confocal microscope, which selects a spot of ~1 µm diameter. The laser-spot diameter amounts to approximately 2 µm. The evaluation procedure follows that of Ref. [6].

**Visualization.** The schematic depiction of the WSe$_2$ monolayer in Fig. 1a is based on crystallographic data provided by the *Materials Project* [47] and drawn by the tool *Mercury* [48].

Acknowledgement

The authors acknowledge financial support by the German Research Foundation (DFG: SFB1083, RA2841/5-1), by the Philipps-Universität Marburg, and the German Academic Exchange Service (DAAD). Synthesis of WSe$_2$ and heterostructure assembly are supported by the NSF MRSEC program through Columbia in the Center for Precision Assembly of Superstratic and Superatomic Solids (DMR-1420634).

Authors' contributions

A.R.-I. conceived the experiment and initiated the study on 2D-exciton-dispersion measurements in 2015. A.R.-I. guided the joint work together with J.C.H. High-quality WSe$_2$ synthesis and heterostructure assembly were achieved by S.S.E., D.A.R., K.B. and J.C.H. The setup was established by L.M.S. and A.R.-I., and the structures measured by L.M.S. The results were interpreted, discussed and summarized in a manuscript by L.M.S. and A.R.-I. with the support of all co-authors.

**Corresponding author**

Arash Rahimi-Iman: a.r-i@physik.uni-marburg.de

James C. Hone: jh2228@columbia.edu

**Authors' statement/Competing interests**

The authors declare no conflict of interest

**Additional information**

Supplementary Information accompanies this paper

## Supporting information

# Optical Measurement of Pseudo-Spin Texture of the Exciton Fine-Structure in Monolayer WSe$_2$ within the Light Cone

Lorenz Maximilian Schneider[1], Shanece S. Esdaille[2], Daniel A. Rhodes[2], Katayun Barmak[3], James C. Hone[2,*], and Arash Rahimi-Iman[1,*]

[1]Faculty of Physics and Materials Sciences Center, Philipps-Universität Marburg, Marburg, 35032, Germany

[2]Department of Mechanical Engineering, Columbia University, New York, NY 10027, USA

[3]Department of Applied Physics and Applied Mathematics, Columbia University, New York, NY 10027, USA

**S1. Reflection-Contrast Overview Spectrum**

To clarify the energetics and to support an overview on the energy–momentum dispersion within the measurable light cone using the Fourier-space spectroscopy method, a white-light reflection contrast measurement over a wider spectral range has been carried out which complements the PL measurements discussed in the main document (see Ref. [6] with regard to initial dispersion characterizations in PL and reflection contrast for the 1s resonance without polarization considerations). First of all, this diagram shown in **Figure SI.1** reveals the actual energies of the relevant resonances, labelled $A_{1s}$, $A_{2s}$, $B_{1s}$, and A-continuum in the angle-integrated line-spectrum of **Figure SI.1a**. This already provides information about the excitation detuning with respect to individual modes. Coloured arrows indicate the excitation energies used in this work (cf. **Figure 1**). For comparison, the angle-integrated PL spectrum of the quasi-resonantly excited 1s state is displayed as a blue curve with one main peak and a shoulder attributed to photon scattering processes transferring dark states into the light cone.

In this overview Fourier-space spectrum, a measurable dispersion can be found not only for the 1s state, but also for the 2s state, although barely visible (see **Figure SI.1b**). Close-up views on the 1s and 2s dispersions are provided in **Figures SI.1c and d**. Nevertheless, the automatic extraction routine (data indicated by dots projected onto the spectrum) still allows one to extract a 2s dispersion that is about one order of magnitude lower in curvature than for the 1s-state. Correspondingly, the effective mass approximation would lead to an order of mass heavier particle in the 2s state. In terms of polariton modes—provided that a zero crossing of the permittivity takes place, which happens only for systems with very narrow homogeneous linewidth—this shallow dispersion would corresponds to a smaller hybridization with the propagating light field due to a smaller oscillator strength (much smaller LT splitting than for the 1s state).

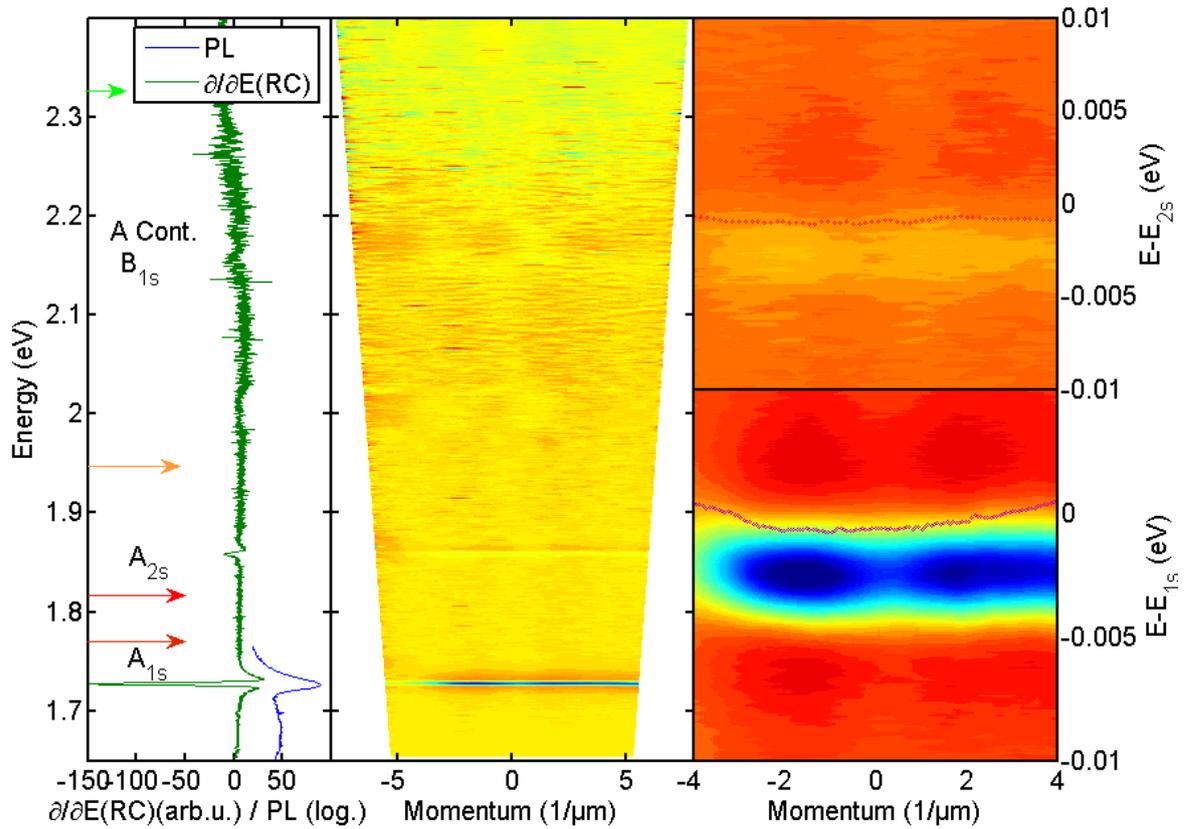

**Figure SI.1 | Overview reflection contrast spectrum.** (a) Comparison of the derivative of the measured reflection contrast (integrated over all measured angles, signal in arbitrary units) with an angle-integrated PL signal (intensity in logarithmic scale, arb. u.) from the same flake. (b) Incidence-angle/momentum dependence of the derivative of the measured reflection contrast. (c & d) Magnified energy–momentum contour maps of the 1s and 2s state with dotted dispersion indicating the exciton energies extracted.

## S2. Determination of the Energy Splitting of the Exciton Branches

As the expected splitting of the exciton branches $\Omega_{ex}(k)$ (approximately 0 – 2 meV) in the detectable $k_{||}$ range is less than the linewidth of the resonances which are quasi superposed on each other, the first-derivative spectrum is used to estimate a splitting value manually as it enables clearer identification of the resonance positions. An exemplary first-derivative PL Fourier-space spectrum is shown in Fig. SI.2. Note that the zero-crossing, i.e. the 1$^{st}$ resonance, which is white in the contour plot and overlaps with the lower flat dashed line indicating the lower branch as a guide to the eye, is not parallel with the position of the strongest negative slope especially for larger momenta. The onset of this slope in this plot, i.e. the shoulder of the strongly momentum-dependent 2$^{nd}$ resonance is marked with the upper angled dashed line representing the clear upwards trend of that feature. The difference between both lines thus can be used as a splitting estimate. A readout error for the splitting value of 0.5 meV is assumed during this extraction method.

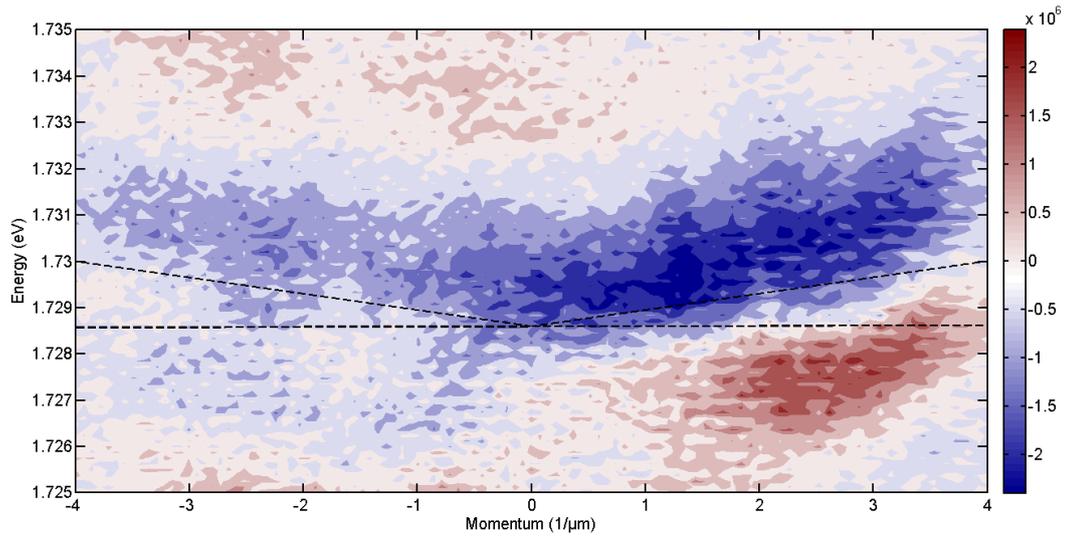

**Figure SI.2 | Derivative of the PL Fourier-space spectrum for quasi-resonant excitation.** The plot shows an exemplary derivative for the smallest pump detuning. The dashed lines serve as guides to the eye. The lower flat line indicates the lower-branch position which is hardly curved. In contrast, the high-slope features of the derivative show a clear momentum-dependent dispersion with minimum at zero in-plane momentum. The upper dashed line is aligned with the onset of the strong negative slope.